# Predicting local and non-local effects of resources on animal space use using a mechanistic step selection model

Jonathan R. Potts[1]*, Guillaume Bastille-Rousseau[2], Dennis L. Murray[2], James A. Schaefer[2] and Mark A. Lewis[1,3]

[1]*Centre for Mathematical Biology, Department of Mathematical and Statistical Sciences, University of Alberta, Edmonton, AB T6G 1G1, Canada;* [2]*Environmental and Life Sciences Graduate Program, Trent University, Peterborough, ON K9J 7B8, Canada; and* [3]*Department of Biological Sciences, University of Alberta, Edmonton, AB T6G 2E9, Canada*

### Summary

1. Predicting space use patterns of animals from their interactions with the environment is fundamental for understanding the effect of habitat changes on ecosystem functioning. Recent attempts to address this problem have sought to unify resource selection analysis, where animal space use is derived from available habitat quality, and mechanistic movement models, where detailed movement processes of an animal are used to predict its emergent utilization distribution. Such models bias the animal's movement towards patches that are easily available and resource-rich, and the result is a predicted probability density at a given position being a function of the habitat quality at that position. However, in reality, the probability that an animal will use a patch of the terrain tends to be a function of the resource quality in both that patch and the surrounding habitat.

2. We propose a mechanistic model where this non-local effect of resources naturally emerges from the local movement processes, by taking into account the relative utility of both the habitat where the animal currently resides and that of where it is moving. We give statistical techniques to parametrize the model from location data and demonstrate application of these techniques to GPS location data of caribou (*Rangifer tarandus*) in Newfoundland.

3. Steady-state animal probability distributions arising from the model have complex patterns that cannot be expressed simply as a function of the local quality of the habitat. In particular, large areas of good habitat are used more intensively than smaller patches of equal quality habitat, whereas isolated patches are used less frequently. Both of these are real aspects of animal space use missing from previous mechanistic resource selection models.

4. Whilst we focus on habitats in this study, our modelling framework can be readily used with any environmental covariates and therefore represents a unification of mechanistic modelling and step selection approaches to understanding animal space use.

## Introduction

Uncovering how space use patterns emerge from animal movement is key to understanding a wide range of ecological phenomena, from disease spread (Kenkre *et al.* 2007; Giuggioli, Pérez-Becker & Sanders 2013) to predator–prey dynamics (Lewis & Murray 1993), conservation biology (Beier 1993) to population density (Grant & Kramer 1990). The desire for animals to find the resources they need to survive and reproduce is a fundamental driver of movement in a variety of animal populations (McIntyre & Wiens 1999; Fortin *et al.* 2003; Breed *et al.* 2009; Houston, Higginson & McNamara 2011). Consequently, many theoretical efforts to understand space use have focused on how animals find and select resources from those available to them (Börger, Dalziel & Fryxell 2008).

Resource selection function (RSF) analysis (Manly *et al.* 2002) is one class of techniques that has been used to address this problem, ever since the seminal paper of Manly (1974). This approach posits that the probability of an animal relocating to a particular patch is a function of both the availability and quality of the resources in the patch. More recently, the studies of Fortin *et al.* (2005) and Rhodes *et al.* (2005) introduced the idea of integrating the RSF with the movement processes of animals, building on the work of Arthur *et al.* (1996). Fortin *et al.* (2005) coined the notion of a step selection function (SSF), where the selection of resources, or other environmental features, directly affects the distance and turning angle of each step. Meanwhile, Rhodes *et al.* (2005) constructed a function for the movement of an animal from one location to the next based on an RSF. These approaches were unified and extended by Forester, Im & Rathouz (2009), who constructed a function for the movement of animals between successive turns based on the previous two positions of the animal, together with the various environmental covariates that affect its movement.

*Correspondence author. E-mail: jrpotts@ualberta.ca

Parallel to these developments, mechanistic models have been constructed that describe the detailed underlying movement processes of animals and derive from them the resulting utilization distribution of animal locations (Moorcroft & Lewis 2006). For many years, this approach developed more or less independently from the RSF methods. However, the study of Moorcroft & Barnett (2008) made inroads into unifying the two theories, by constructing a mechanistic movement kernel based on an RSF and deriving from that the probability distribution of the animal. This showed, for the first time, how RSF analysis could be used to link analytically the movement processes of animals with the emergent features of its space use.

In the model of Moorcroft & Barnett (2008), the probability of an animal being in a particular location turns out to be a function of the quality of resources at that location. Whilst this is a sensible first approximation, one of the consequences of this model is that animals are just as likely to be found in small isolated patches of good habitat than within large contiguous areas of habitat of equal quality. In reality, both isolation and size of patches are key drivers of space use in many animal populations (Andrén 1994; Hill, Thomas & Lewis 1996; Bender, Contreras & Fahrig 1998). Ideally, mechanistic models that predict space use accurately should give rise to utilization distributions where occupation probability is positively correlated with patch size and negatively correlated with isolation.

In this study, we describe a novel mechanistic model of animal movement where the resulting utilization distributions include both of these features. We also demonstrate how to parametrize the model from location data, using herds of caribou (*Rangifer tarandus*) in Newfoundland as an example. There are about 14 major caribou herds on Newfoundland Island. Most herds exhibit semimigratory behaviour involving philopatric movements, with females moving every year to traditional calving grounds during spring and summer (Mahoney & Schaefer 2002). The data we use are of movement within these calving grounds.

Our model is based on an SSF, from which we derive a mechanistic master equation, allowing us to compute numerically the steady-state probability distribution of the animal positions, thus relating quantitatively the movement processes to the emergent space use patterns. Relative intensity of space use in a given place is a function of different movement responses that involve both variation in mean displacements within habitats and preferential movement directions towards preferred areas (Bastille-Rousseau, Fortin & Dussault 2010). Whilst resource selection analysis does not disentangle explicitly the mechanisms involved (Bastille-Rousseau, Fortin & Dussault 2010), most mechanistic models do not consider how animals move selectively from one specific resource to another.

Our approach addresses this by modelling the movement decision based not on the absolute quality of the habitat to where the animal might move, but the relative quality of this habitat compared with the habitat where the animal is currently positioned. Studies of optimal foraging strategies in mice (Morris & Davidson 2003) demonstrate that short-term movement decisions of individuals are grounded in the relative fitness associated with the habitats between which they are moving. Constructing mechanistic movement models that are based on behavioural decisions arising from underlying evolutionary forces is important if we wish to understand not just *how* space patterns form but *why*. Though the results of Morris & Davidson (2003) are based on mouse populations, their underpinning in the general theory of natural selection suggests that these ideas may well extend to other taxa. By grounding the SSF in ideas from optimal foraging theory, one would expect the model outcomes to be closer to those observed in real ecological systems.

Indeed, our simple change in the formulation of the step selection mechanism causes dramatic changes in the utilization distribution, as the effect of resources on the resulting position distribution of the animals propagates through the landscape via their movement processes. In particular, the effects of patch size and isolation on animal utilization become apparent, which are not present in previous mechanistic models. We believe that this modelling framework will prove useful in building simple yet accurate predictive models of the underlying determinants of complex space use patterns, that account for both the non-local as well as the local effects of environmental features.

## Materials and methods

### THE MASTER EQUATION

The master equation (ME) is the key building block in linking individual processes to population patterns. It is defined to be an equation built from individual movement decisions that gives the probability density at some time $t + \Delta t$ as a function of the probability density at time $t$, where $\Delta t$ is some fixed time interval, for example the time between animal location fixes. As such, it is an example of a one-step Markov process.

The ME for our model is based on a step selection framework introduced by Fortin *et al.* (2005) and extended by Forester, Im & Rathouz (2009), which gives the probability of moving from one location to the next in a given time interval (i.e. a step). Whilst we use a correlated random walk framework similar to Forester, Im & Rathouz (2009), we find it convenient to reformulate the step selection function (SSF) as follows

$$f(x|y, \theta_0) = \frac{\Phi(x|y, \theta_0)\mathcal{W}(x, y, E)}{\int_\Omega dx' \Phi(x'|y, \theta_0)\mathcal{W}(x', y, E)}, \quad \text{eqn 1}$$

where $f(x|y, \theta_0)$ is the probability of finding an animal at position $x$, having travelled from $y$ in the previous step, given that it arrived at $y$ on a bearing of $\theta_0$ (bearings are measured in an anti-clockwise direction from the right-hand half of the horizontal axis), $\Phi(x|y, \theta_0)$ is the probability of being at $x$ in the absence of habitat selection, given that the animal was previously at $y$ and had arrived there on a bearing of $\theta_0$ and $E$ contains details about the environment that we wish to model. In terms of classical resource selection, $\Phi(x|y, \theta_0)$ can be thought of as a function detailing how available $x$ is to the animal. Typically, it will decay the further $x$ is away from $y$ so that distant places are less available than nearby areas.

For example, in Fortin *et al.* (2005), $E$ contains the distribution of forest types in the study area, information about predator positions, snow abundance, topography and road locations. There, $\mathcal{W}(x, y, E)$ is a function of $x$, $y$ and $\varepsilon$ that measures features such as the proportion of the line segment from $y$ to $x$ containing conifer forest, the minimum distance of this line segment to a road, and various other important environmental aspects that affect the animals' movement (Fortin *et al.* 2005).

The area to which the animal is confined is denoted by $\Omega$. This may be a geographical limitation of the movement, such a small island, or confinement to a home range or territory. For certain populations, the latter may not be stationary over time (Potts, Harris & Giuggioli 2013), requiring $\Omega$ to be replaced by a time-dependent function $\Omega(t)$. The size and shape of $\Omega(t)$ may in turn depend upon the past positions of animals in neighbouring territories. However, for the purposes of this study, we will assume $\Omega$ is constant.

The denominator in eqn (1) simply ensures that the function $f(x|y, \theta_0)$ is a probability density function; that is, it integrates to 1 with respect to $x$. The variable $x'$ is a dummy variable of integration, used to distinguish positions in the domain of integration from $x$, the position to which the animal is moving.

For this study, we divide $\Omega$ into habitat types $H_i$, $1 \leq i \leq M$. The set of all habitat types is denoted by $\mathcal{H}$, so that $H_i \in \mathcal{H}$. We construct an SSF that is based on the habitat at both the beginning of the step and the end of the step, with the ultimate aim to understand numerically how this affects the resulting animal space use distribution. This is an aspect missing from current work on SSFs or RSFs, with the exception of the simpler model in Moorcroft & Barnett (2008). We use the non-negative number $W(H_i, H_j)$ to denote the tendency for the animal to move from habitat $H_j$ to $H_i$, depending on how preferable the habitat $H_i$ is compared with $H_j$. If $W(H_i, H_j) > 1$ then $H_i$ is more preferable than $H_j$, whereas $W(H_i, H_j) < 1$ means $H_j$ is more preferable than $H_i$. We denote by $H(x)$ the habitat at position $x$. The functional form of our SSF is then

$$f(x|y, \theta_0) = \frac{\Phi(x|y, \theta_0) W[H(x), H(y)]}{\int_\Omega dx' \Phi(x'|y, \theta_0) W[H(x'), H(y)]}. \qquad \text{eqn 2}$$

Notice that $W[H(x), H(y)]$ can be written as $\mathcal{W}(x, y, H)$ to put it in the form given in eqn (1). However, we choose the former notation as we believe it to be more instructive for our particular function.

Equation (2) gives rise to the following ME for the probability density function $u(x, \theta, t + \Delta t)$ of the animal being at $x$ at time $t + \Delta t$ having travelled there on a bearing of $\theta$

$$u(x, \theta, t + \Delta t) = \int_{-\pi}^{\pi} d\theta_0 \int_0^{s_{max}} ds \frac{\Phi(x|y_\theta(s), \theta_0) W[H(x), H(y_\theta(s))]}{\int_\Omega dx' \Phi(x'|y_\theta(s), \theta_0) W[H(x'), H(y_\theta(s))]} u(y_\theta(s), \theta_0, t), \qquad \text{eqn 3}$$

where $y_\theta(s)$ describes the locus of points $y$ upon which the animal could approach $x = (x_1, x_2)$ at bearing $\theta$, that is, $y_\theta(s) = (x_1 + \cos(\theta + \pi)s, x_2 + \sin(\theta + \pi)s)$, with $s$ denoting the distance between $y_\theta(s)$ and $x$. Here, $s_{max}$ is the distance along this line from $x$ to the boundary of $\Omega$ and so gives the upper endpoint of integration. Though eqn (3) may look formidable, in practice, it is simple to implement by discretizing space (see Appendix S1).

## DATA COLLECTION METHODS

Since 2006, more than 200 caribou were captured during winter and fitted with GPS collars that acquired locations every two hours. We focus our study on 140 caribou followed between 2006 and 2012 and limit analysis to six distinct herds, which had sufficient amounts of individuals and monitoring. The other caribou were ignored since there were only a small number per herd. We limit our movement analysis to the critical, non-migratory period of calving and post-calving (May 1 to September 1), which gives us more than 300 000 position fixes at two-hourly intervals. Every location is given a characterization based on the habitat it falls into, using a reclassified Landsat TM imagery (Wulder *et al.* 2008). Collar equipment use and capture methods are consistent with American Society of Mammalogists guidelines (Gannon & Sikes 2007). On rare occasions, a position fix failed to be recorded (0·997% of fixes). In each of these cases, we split the data at that point, so that we only considered steps that were two hours long. Data required for repeating this study are available from the Dryad Digital Repository: http://doi.org/10.5061/dryad.1d60p.

## MODELLING CARIBOU MOVEMENT

Variations in habitat type can affect animal behaviour, in particular their step length and turning angle distributions $\Phi(x|y, \theta_0)$ (Moorcroft & Lewis 2006). Therefore, to capture correctly the effect of movement processes on the space use distribution, it is necessary to split $\Phi(x|y, \theta_0)$ into a sum of functions, one for each habitat type, as follows

$$\Phi(x|y, \theta_0) = \sum_{h \in \mathcal{H}} I(y, h) \Phi_h(x|y, \theta_0), \qquad \text{eqn 4}$$

where $I(y, h)$ is an indicator function taking the value 1 if $H(y) = h$ and 0 otherwise, and $\mathcal{H}$ is the set of all habitat types available to the animal. Notice that this only depends on the habitat where the animal currently resides (position $y$) so that $\Phi(x|y, \theta_0)$ is independent of the selection of the next habitat the animal is to move to (at position $x$).

$\Phi(x|y, \theta_0)$ contains information about both the step length, that is, the distance travelled in successive relocations and the turning angle between successive steps. Whilst in general, the distribution of these two aspects of movement may depend on one another, a linear–circular correlation test between the step length and turning angle distributions for the caribou data has $R^2 = 0·027$, suggesting the two distributions are not tightly correlated. Therefore, we assume that they are independent, so that

$$\Phi_h(x|y, \theta_0) = V_h[\psi(x, y, \theta_0)] \rho_h(|x - y|), \qquad \text{eqn 5}$$

where $V_h(\phi)$ is the turning angle distribution for habitat $h$, $\psi(x, y, \theta_0)$ calculates the turning angle for an animal that has just travelled to $y$ on a bearing of $\theta_0$ and turns to move in a straight line towards $x$, and $\rho_h(r)$ is the step length distribution for habitat $h$. For the step lengths, we tried fitting exponential and Weibull distributions and found the Weibull distribution to give the best fit, using a likelihood ratio test. This has the following form

$$\rho_h(x|a, b) = \frac{a}{b} \left(\frac{x}{b}\right)^{a-1} \exp\left[-\left(\frac{x}{b}\right)^a\right]. \qquad \text{eqn 6}$$

For the turning angles, we tried fitting both univariate and bivariate von Mises distributions (McKenzie *et al.* 2012) using a likelihood ratio test. The latter was tested because caribou may make use of linear features, such as paths, and it turned out to be the better of the two. It has the following form

$$V_h(\phi|k_1, k_2) = \frac{\exp[k_1 \cos(\phi)]}{4\pi I_0(k_1)} + \frac{\exp[k_2 \cos(\phi - \pi)]}{4\pi I_0(k_2)}, \quad \text{eqn 7}$$

where $I_0(x)$ is the zeroth order modified Bessel function of the first kind. The first summand in eqn (7) represents the tendency for the caribou to continue in a similar direction. The second summand is due to a bias in the data towards the caribou performing 180° turns between successive steps. Part of the measured bias towards 180° turns can be due to errors in GPS measurements (Hurford 2009), so to rule this cause out, we removed locations where caribou moved <25 m, following the methods detailed in Hurford (2009). This amounted to 19·8% of the fixes. Notice that we only removed such data for the analysis of turning angles, not for the step lengths or the resource weighting function $W(H_i, H_j)$. We found that the bivariate von Mises distribution provides a better fit than the univariate von Mises distribution for the turning angles.

### PARAMETRIZING THE MASTER EQUATION FROM LOCATION DATA

We estimated the parameters $a$, $b$, $k_1$ and $k_2$ for the function $\Phi(x|y, \theta_0)$ using the maximum-likelihood method, with the Nelder–Mead simplex algorithm (Lagarias *et al.* 1998). To calculate the resource weighting function $W(H_i, H_j)$, we wish to capture the probability $P(H_i|y, \theta_0, w_{ij})$ of an animal moving into habitat type $H_i$, given its present position $y$, trajectory $\theta_0$ and weights $w_{ij} = W(H_i, H_j)$ (see Fig. 1). In other words, we aim to maximize the likelihood function

$$\prod_{n=2}^{N} P(H(x_n)|x_{n-1}, \theta_{n-1}, w_{ij}), \quad \text{eqn 8}$$

where $x_1, x_2, \ldots, x_N$ and $\theta_1, \theta_2, \ldots, \theta_N$ are the data on the animal's position and bearings, respectively, and

$$P(H(x_n)|x_{n-1}, \theta_{n-1}, w_{ij}) = \frac{\int_{\Omega_i} dx \Phi(x|x_{n-1}, \theta_{n-1}) W[H(x), H(x_{n-1})]}{\int_{\Omega} dx \Phi(x|x_{n-1}, \theta_{n-1}) W[H(x), H(x_{n-1})]}, \quad \text{eqn 9}$$

where $\Omega_i = \{x \in \Omega | H(x) = H_i\}$.

Whilst it is, in principle, possible to calculate the maximum of the likelihood function (eqn 8) by numerically evaluating the integrals in eqn (9) for each data point, this is highly computationally intensive if the data set is large, which is often the case with GPS telemetry data. We instead choose a more efficient method that makes use of a Monte Carlo sampling procedure. For each $n \in \{2, 3, \ldots, N\}$, where $\{x_1, \ldots, x_N\}$ is the set of animal locations, we sample $M = 100$ times from $\Phi(x|x_{n-1}, \theta_{n-1})$ to give a set $S_n$ of possible next animal positions, disregarding the biasing effect that resources have on the movement. The reason for using $M = 100$ is to reduce computational time for analysing our large data set (>300 000 steps), and by examining a small subset of the data using $M = 1000$, we obtain similar results to $M = 100$. We then use the approximation $\int_{\Omega_i} dx \Phi(x|x_{n-1}, \theta_{n-1}) \approx |\{s \in S_n | H(s) = H_i\}|/|S_n|$ to give

$$P(H(x_n)|x_{n-1}, \theta_{n-1}, w_{ij}) = \frac{\int_{\Omega_i} dx \Phi(x|x_{n-1}, \theta_{n-1}) W[H(x), H(x_{n-1})]}{\sum_j \int_{\Omega_j} dx \Phi(x|x_{n-1}, \theta_{n-1}) W[H(x), H(x_{n-1})]}$$

$$\approx \frac{W[H_i, H(x_{n-1})]|\{s \in S_n | H(s) = H_i\}|}{\sum_{s \in S_n} W[H(s), H(x_{n-1})]}, \quad \text{eqn 10}$$

where $|S|$ denotes the number of elements in a set $S$, so that the likelihood function is

$$L(w_{ij}) = \prod_{n=2}^{N} \frac{W[H_i, H(x_{n-1})]|\{s \in S_n | H(s) = H_i\}|}{\sum_{s \in S_n} W[H(s), H(x_{n-1})]}. \quad \text{eqn 11}$$

To maximize eqn (11) efficiently, we split it into several likelihood functions $L_j$, one for each habitat type $H_j$

$$L_j(w_{ij}) = \prod_{n-1 \in Q_j} \frac{w_{ij}|\{s \in S_n | H(s) = H_i\}|}{\sum_{s \in S_n} W[H(s), H_j]}, \quad \text{eqn 12}$$

where $Q_j$ is the set of indices $m$ such that $H(x_m) = H_j$. For each $j$, we maximize the corresponding likelihood function (eqn 12) independently of the others, whilst ensuring that $w_{jj} = 1$, using the Nelder–Mead simplex algorithm (Lagarias *et al.* 1998), as implemented in the Python `maximize()` function from the SciPy library (Jones, Oliphant & Peterson 2001). The likelihood function for the entire data set is simply the product $L(w_{ij}) = \prod_{j=1}^{|\mathcal{H}|} L_j(w_{ij})$, where $|\mathcal{H}|$ is the number of habitat types. To obtain error bars for the weights, we bootstrapped the set of steps 100 times and calculated the maximum likelihood parameter values for each. Error bars are standard deviations of the results.

### NUMERICAL INVESTIGATION OF THE MODEL

To investigate the model, we constructed artificial resource landscapes on a 50 by 50 square lattice, where the lattice spacing is 200 m. We used the weighting function $W(H_i, H_j)$, step length and turning angle distributions found by fitting to the caribou data, as described in the previous subsection. We computed the steady-state position distribution numerically on this lattice by iterating the master equation (eqn 3) through time until $|u(x, \theta, t + \delta t) - u(x, \theta, t)| < 10^{-8}$ for every value of $x$ and $\theta$.

To understand how patch size and isolation affect the steady-state probability distribution, we used artificial landscapes where the left-hand half is wetland habitat and all of the right-hand half is coniferous dense forest, except for a small square of wetland, which we call *the patch*. The reason for having the left-hand half of the terrain as good-quality wetland habitat is that animals need to be given a choice between a small patch and a large contiguous area of good habitat. If the terrain contains just a single good patch on its own in the middle of poor habitat, then the animals will choose the good patch with high probability even if it is small and isolated, as it is the best option available. We also used the same step length and turning angle distribution in both habitats, so as to isolate the effect of the weighting function on space use patterns.

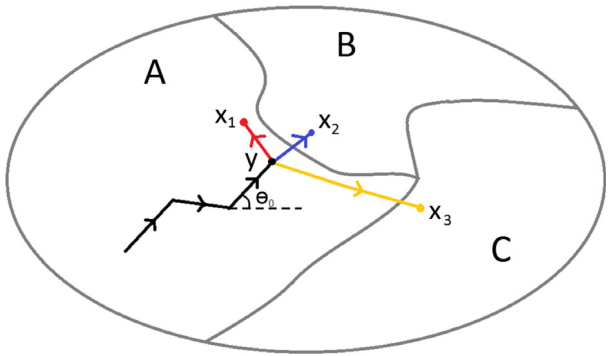

**Fig. 1.** Schematic representation of the movement model. The animal represented here has moved to point $y$ on the trajectory given by the black lines in an environment with three resource types: A, B and C. Suppose that C is the most preferable habitat for the animal, followed by A, with B being resource poor. Three of the many possible next steps for the animal are to $x_1$, $x_2$ or $x_3$. In the absence of a resource response, and assuming that the animal is a correlated walker with a step length distribution that decays with increasing distance, the most likely move would be to $x_2$ in patch B. However, due to the poor quality of patch B, the animal may instead decide to take a sharp left turn to stay in patch A (represented by a move to $x_2$) or even to take a sharp right turn and move the longer distance in order to end in the highest-quality patch C (represented by a move to $x_3$).

To investigate the patch size effect, we placed the patch 1·2 km to the right of the centre of the landscape and halfway up. We varied the patch size from 0·16 to 7·84 km$^2$. To make sure that the overall amount of wetland and forest was the same in each artificial landscape, we replaced a strip of wetland on the left-hand side of the landscape with coniferous dense forest, ensuring that the area of the strip was the same as that of the patch. To examine the isolation effect, the patch was placed at differing distances to the right of the landscape centre, between 0·4 and 3·2 km, and the patch size was kept constant at 1·6 km$^2$.

## Results

We identified five different habitat types within the landscape: wetland (WL), barren (B), dense coniferous forest (CD), open coniferous forest (CO) and other (O). The O category consists of water and other non-abundant resources, such as byroids, herbs and broadleaf. For all of these habitats, the bivariate von Mises distribution for the turning angles and the Weibull distribution for the step lengths were good fits to the data (Fig. 2, Table 1).

The best-fit parameters for the weighting function $W(H_i, H_j)$ are shown in Table 2. This table suggests that WL should be the most favourable habitat type, since the weight given to moving there from other habitats is always >1. B is close behind, being preferable to all other habitats except wetland. CO is a middling habitat, with half the weights of moving there being >1 and the other half less. CD appears to be notably less preferable to these first three, with O being the least favourable of all categories.

**Table 1.** Step lengths and turning angles for the caribou data

| Resource type | $a$ | $b$ (m) | $k_1$ | $k_2$ |
| --- | --- | --- | --- | --- |
| Barren (B) | 0·754 | 346 | 1·498 | 0·490 |
| Wetland (WL) | 0·688 | 289 | 1·292 | 0·573 |
| Coniferous dense (CD) | 0·677 | 189 | 0·762 | 0·733 |
| Coniferous open (CO) | 0·677 | 214 | 0·933 | 0·673 |
| Others (O) | 0·604 | 212 | 1·028 | 1·057 |

Parameter values for the caribou step length and turning angle distributions, given in eqns (6) and (7), respectively. There is one step length distributions for each habitat type, depending on the habitat at the start of the step. Each turning angle distribution depends upon the habitat at the point where the animal makes the turn. The parameter $b$ is measured in metres and the other parameters are dimensionless. To measure the turning angle distributions, we removed any steps of <25 m, as recommended by Hurford (2009). This required removing 8·4%, 22·6%, 55·3%, 8·9% and 4·7% of the angles in the B, WL, CO, CD and O habitats, respectively.

When a small WL patch is placed in the midst of an area of CD, in the simulated environment described in the methods section, the average space use per unit area increases with the size of the patch but decreases with isolation (Fig. 3), showing the effect of the weighting function on the emergent space use patterns. However, variations in step length and turning angle distributions also play an important role. The further an animal moves between fixes, the faster it is moving on average, which affects the animal's space use distribution. Previous mechanistic models (Moorcroft, Lewis & Crabtree 2006; Moorcroft & Lewis 2006) have shown that some animals, for example coyote (*Canis latrans*) (Laundré & Keller 1981), will decrease their speed of movement in more favourable habitats and that this causes them to be observed with higher probability in better habitats than worse ones. However, in certain circumstances, some species, for example elk (*Cervus elaphus*) (Anderson, Forester & Turner 2008) and black bears (*Ursus americanus*) (Bastille-Rousseau *et al.* 2011), do not appear to slow down in preferred habitats.

Similar to these latter examples, the caribou in our study move fastest in habitats B and WL (Table 1), most likely because these habitats are open and offer relatively few obstructions to movement for a large and long-legged animal such as caribou. However, B and WL appear from the resource weightings (Table 2) to be preferable to the other three habitats. Since the faster movement in B and WL would cause the caribou to spend relatively less time in these habitats than would be expected if all the step length distributions were equal, we have competing effects between fidelity to these habitats due to the tendency to move into these habitats from others and lower space use caused by faster moving within B and WL.

To examine these competing effects, we computed numerically the steady state of the ME (eqn 3) in an artificial landscape consisting of just WL and CD habitat types (Fig. 4a). These are chosen because animals move fast in WL (Table 1) but it is the most preferential habitat according to the weighting function (Table 2), whereas animals move slowly in CD but do not choose this habitat preferentially over WL, B or CO. When the step length distributions for both habitats are the same (Fig. 4b), there is a clear preference for WL. In addition to this, the probability density is highest in the largest contiguous WL area, towards the bottom-left than in the other, smaller patches. The smallest patches of WL, in the top-left and top-right, show the lowest probability density of all the WL patches. This is a feature of space use that does not emerge in the mechanistic resource selection model of Moorcroft & Barnett (2008). In that model, the space use at any point is a function of the resource quality at that point, so that the probability density would be of the same magnitude in all the WL patches. Here, the preference of animals for large, contiguous patches of high-quality habitat emerges naturally from the underlying movement processes.

When we solve the steady state of the ME (eqn 3) in the same landscape, but this time with different step length distributions in different habitat types, as given in Table 1, very different space use patterns emerge (Fig. 4c). CD is much more

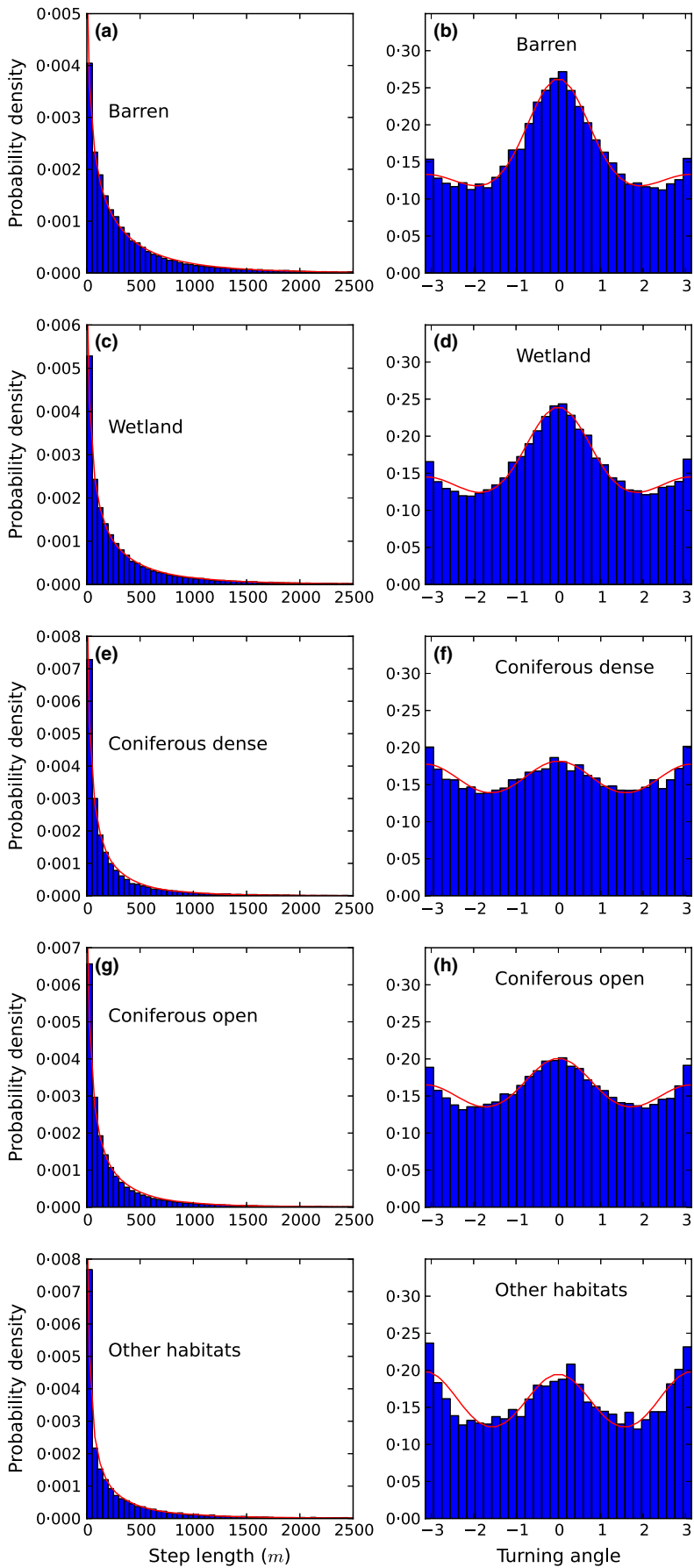

**Fig. 2.** Step length and turning angle distributions for the various habitat types. Blue bars show the probability densities from the caribou data and red lines the best fit curves for eqn (6) (left-hand panels) or eqn (7) (right-hand panels). Parameter values for these curves are given in Table 1. The left-hand charts are step length distributions and the right-hand are turning angle distributions. The habitat types from top to bottom are Barren, Wetland, Coniferous dense, Coniferous open, and Other habitats.

**Table 2.** Resource weights for the caribou data

| From | to B | to WL | to CD | to CO | to O |
| --- | --- | --- | --- | --- | --- |
| Barren (B) | 1·00 | 1·05 ± 0·01 | 0·63 ± 0·02 | 0·89 ± 0·01 | 0·41 ± 0·01 |
| Wetland (WL) | 0·95 ± 0·01 | 1·00 | 0·64 ± 0·01 | 0·93 ± 0·01 | 0·38 ± 0·01 |
| Coniferous dense (CD) | 1·17 ± 0·04 | 1·08 ± 0·02 | 1·00 | 1·06 ± 0·01 | 0·35 ± 0·01 |
| Coniferous open (CO) | 1·07 ± 0·01 | 1·07 ± 0·01 | 0·82 ± 0·01 | 1·00 | 0·29 ± 0·01 |
| Others (O) | 1·65 ± 0·05 | 1·64 ± 0·05 | 0·92 ± 0·05 | 1·38 ± 0·04 | 1·00 |

The weighting $W(H_i, H_j)$ given to travelling from one habitat $H_j$ to another $H_i$, calculated from the caribou data. $W(H_i, H_j) > 1$ means that $H_i$ is preferable to $H_j$, whereas movement from a more preferable habitat to less means $W(H_i, H_j) < 1$. Consequently, $W(H_j, H_j) = 1$ for any $H_j$. Columns denote the habitat type to which the animal is moving and rows denote the habitat from where the animal came. Each of the non-diagonal entries were significantly different from 1, with $P < 0.0001$, using likelihood ratio test. Error bars are single standard deviations obtained by bootstrapping the data (see 'Materials and Methods').

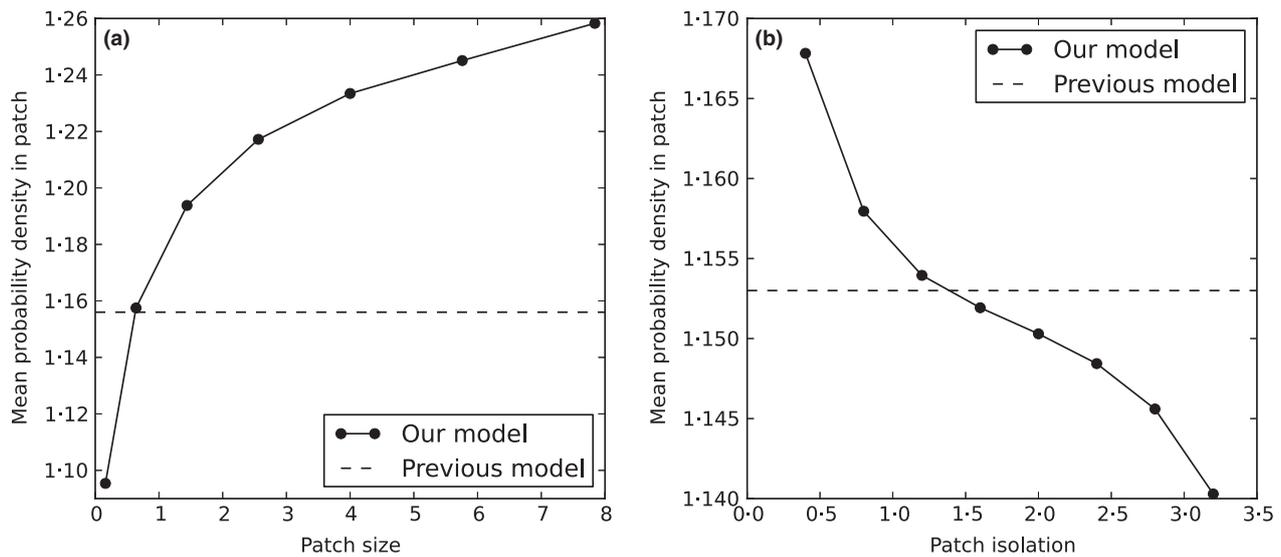

**Fig. 3.** The size and isolation of a patch affect the probability of model animals being found in the patch. Panel (a) shows the average probability density of an animal to be found in a (good quality) wetland patch surrounded by (poor quality) dense coniferous forest, as a function of the size of a patch in $km^2$. Panel (b) shows the same average probability density, this time as a function of the distance of the patch in *km* from a large contiguous area of wetland (see 'Materials and Methods' for details). In both panels, the solid lines show the results of the steady-state solution of the model described in this study. The dashed lines show the results of the steady-state solution of the model described in Moorcroft & Barnett (2008), when the weight of moving to wetland is 1·17 times that of moving to dense coniferous forest. In our model, mean probability density increases with patch size and decreases with patch isolation, whereas neither of these properties of the patch have an effect on the animal probability density in Moorcroft & Barnett's model.

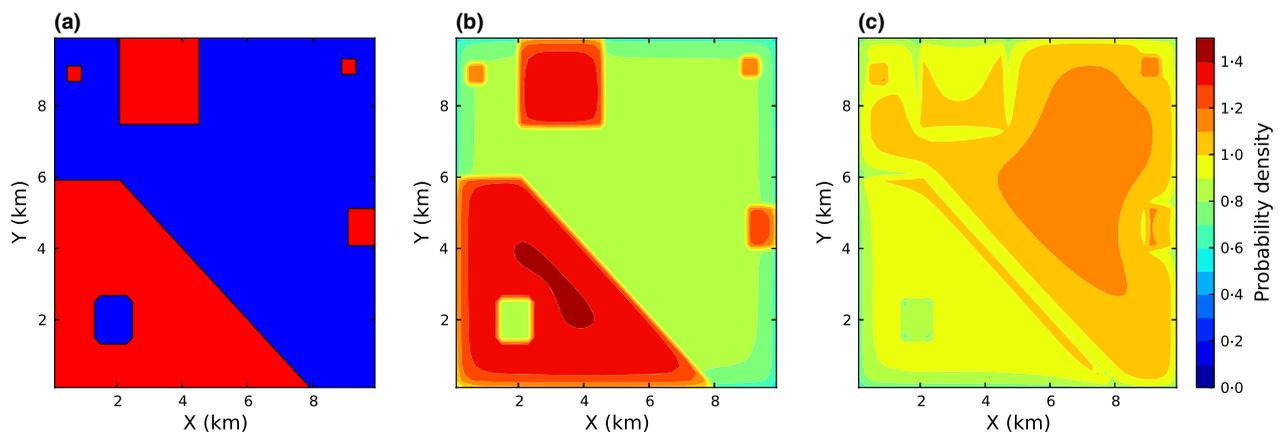

**Fig. 4.** Steady-state solutions of the master equation in an artificial environment. Panel (a) shows the resource distribution, where blue areas are coniferous dense forest (CD) and red sections are wetland (WL). Panel (b) is the steady-state solution of eqn (3) where the resource weights are as in Table 2 but the step length and turning angle distributions are the same for both habitat types. Panel (c) is the steady-state solution of eqn (3) where the resource weights are as in Table 2 and the step length distributions are as in Table 1, that is, different for each habitat type. The turning angle distributions are uniform in both panels (b) and (c). Distances along the *X* and *Y* axes are in kilometres.

preferable in this scenario than that in Fig. 4b. Particularly, the centre of the large contiguous area of CD in the top-right has the highest probability density of the whole landscape. Here, the animals are far away from any WL habitat, and moving slowly, so are less likely to choose preferentially to travel to the WL habitat than stay in the same patch. The only other place where the probability density is as high as in the centre of the large CD area is the small WL patch at the top-right, which is the only patch of WL near the centre of the CD habitat. Conversely, the isolated patch of CD in the bottom-left, surrounded by a large area of WL, is relatively under used, since animals there are always close to the preferable WL habitat, so will tend to move from the CD patch to the surrounding WL area.

Another interesting feature of Fig. 4c occurs along the edge of the large patch of wetland. The probability density at the edge is higher than anywhere else in this wetland patch, owing to the model animals tending to move there if they end up at the neighbouring edge of the forest. A variety of species have been observed to choose preferentially the edge of a good habitat over the interior, for example insects such as large white butterflies (*Pieris brassicae*) (Bergerot *et al.* 2013), mammals such as pygmy tarsiers (*Tarsius pumilus*) (Grow, Gursky & Duma 2013) and reptiles such as black rat snakes (elaphe *obsoleta obsoleta*) (Blouin-Demers & Weatherhead 2001). Our model may go some way to explaining the mechanisms behind this phenomenon.

## Discussion

We have constructed a mechanistic movement model, based on a step selection function (SSF), where the movement is governed by the relative habitat quality between the start and the end of the step. Though simple in concept, this model has complex outcomes that mimic features of space use observed in many animal populations and that are not present in simpler mechanistic resource selection models (Moorcroft & Barnett 2008). As well as patch usage being correlated with local habitat quality, the size and isolation of the patch also affect the space use patterns that emerge from our model. Larger patches of good habitat are more likely to be used than smaller ones of equal quality. Additionally, isolated patches of good habitat inside large areas of bad habitat are less used than patches of similar size and quality that occur near bigger, good-quality patches. Both of these features of space use have been observed in a wide variety of animal populations (Andrén 1994) so it is important for mechanistic models to replicate them in order to make accurate predictions.

We generalized the SSF for a correlated random walk from the version in Forester, Im & Rathouz (2009). The latter is a two-step Markov process, depending upon the position of the animal at the previous two time steps. However, in order to construct a master equation from the SSF (Moorcroft & Barnett 2008), it is convenient to use our one-step Markov process formulation, which depends upon the position and bearing at the previous time-step (eqn 1). We also extended the SSF from Forester, Im & Rathouz (2009) to enable inclusion of information about the whole step, as done in Fortin *et al.* (2005), rather than just the end of the step. A strength of the master equation approach is that it gives the full probability distribution as it evolves over time by solving the equation just once. Since it is not subject to random variation, as is the case when performing stochastic simulations, this obviates the need for simulating multiple realizations or having to determine how many simulations are required to give a full and accurate picture of the model behaviour.

We have explained how to parametrize our model from location data, using herds of caribou in Newfoundland as our test population. This advances the study of Moorcroft & Barnett (2008), which describes purely theoretical results in a mathematically simplified one-dimensional world, and will enable biologists to construct mechanistic step selection models appropriate for their study species. Whilst we have focused on resources in the present paper, our model can be readily extended to include other environmental covariates. Such models could be used to test hypotheses about the mechanisms that cause observed space use patterns to emerge in the population (Moorcroft, Lewis & Crabtree 2006).

Resource selection techniques have been successfully used to uncover the driving factors behind movement decisions for a large variety of populations (Manly *et al.* 2002). However, they cannot, by themselves, relate movement decisions to spatially explicit, population-level patterns of usage in a non-speculative, analytic fashion. Mechanistic models, on the other hand, were developed precisely for this reason: to derive the space use distribution of animals from details of the underlying causal processes (Moorcroft & Lewis 2006). They therefore provide a quantitative link between individual-level and population-level descriptions. This is vital for accurately building and parametrizing models that are often constructed on the population-level, such as those of disease spread and predator–prey dynamics, but whose underlying processes are driven by individual-level movement and interaction events.

Whilst the recent development of SSFs (Fortin *et al.* 2005; Rhodes *et al.* 2005; Forester, Im & Rathouz 2009) has gone a considerable way towards framing resource selection in the context of the animal's movement mechanisms, previous studies have not used the SSF to determine the utilization distribution that the SSF would predict. Here, we demonstrate how to frame an SSF, which can take into account features of the whole step, in such a way as to derive this utilization distribution, via construction of a master equation. This gives a framework for studying how different environmental covariates affect space use patterns. It would therefore be possible to use our techniques to shed light on how the various covariates described in previous step selection studies each affect the way animals use space, thus giving insights into why certain parts of the landscape are used more than others and ultimately helping predict the effect of possible future landscape changes on animal space use.

Behavioural processes such as habitat selection and movement strategies are key components of animal space use, which are considered explicitly by mechanistic models. Notwithstanding the variety in determinants that have been tested in the past, animal behaviour can be far more complex than current mechanistic models consider. Different foraging strategies can lead to an increased use of a given resource; animals can increase time spent in a patch by reducing their rate of movement within a patch or by selectively moving between patch of a specific type (Bastille-Rousseau, Fortin & Dussault 2010) as predicted by optimal foraging theory (Morris & Davidson 2003). The resource weighting function added to our mechanistic model allows explicit representation of such behaviour and may be used to enable researchers to have a better understanding of the foraging strategies animals use. Our resource weighting function also naturally gives rise to real aspects of animal space use such as large areas of good resource being used more intensively than smaller patches, which are important features of animal space use. This occurs by ensuring that the relative quality of the habitat between the start and end of each step is considered, so that the effect of resource quality at a point is propagated through the landscape by the non-local movement decisions of the animal.

However, the weighting function and movement parameters assume that the preference for a given resource or habitat is constant and will not change based on the spatial context that animals are currently in. This assumption may not hold when habitat selection is subject to a functional response; that is, that the selection for a specific attribute is changing with the spatial context (Mysterud & Ims 1998; Hebblewhite & Merrill 2008). Animals living in an area with different availability of resources could display different responses based on feature availability at multiple scales, such as within home range or inter home range (Moreau *et al.* 2012). It may therefore be necessary, when applying our mechanistic step selection model to multiple individuals ranging large areas, to assess first the presence of variation between and within individual behaviour based on habitat availability. Indeed, such an assessment could be made using our modelling framework. To apply the framework to a single animal requires no methodological changes, but simply applying the same techniques to the movement data for a single animal, rather than pooled data as demonstrated here. Based on the scale of the functional response, different parameter estimates could then be obtained for individuals experiencing heterogeneous conditions or for specific areas of the landscape.

The purpose of this paper is not specifically to study caribou behaviour. However, the fact that we have chosen data on this particular species to parametrize our model has opened up various questions about caribou space use that we hope to answer in future work. For example, why do caribou move faster in preferred habitats? It may also be interesting to compare habitat choice over longer temporal scales than two hours. If it can be shown that the choices tend to be made over longer time periods, this would suggest that the animals are using some sort of cognitive map of the environment to determine their movement, rather than simply making choices on a step-by-step basis. Furthermore, temporal differences, such as variations in night-and day-time behaviour, may have an impact on space use, which would be worth investigating in future.

On the more theoretical side, the results of this paper suggest that other choices of parameter may cause the formation of further, qualitatively different spatial patterns. Due to the inherent computational intensiveness of numerical simulations, rigorous and exhaustive analysis of such patterns requires development of an analytic theory of the type of SSFs studied here. Such analysis would also help illuminate the reasons behind the phenomena unveiled by the numerical studies of this paper. We hope to examine these ideas in future work.

The present study deals with the effect of movement processes on space use. However, interactions between animals also have an important effect in many populations, either due to collective grouping phenomena (Couzin *et al.* 2002; Camazine *et al.* 2003) or territorial exclusion (Lewis & Murray 1993; Giuggioli, Potts & Harris 2011). In principle, the latter can be factored in our mechanistic modelling framework by including a term into our SSF (eqn 2) that excludes movement by animals into places recently occupied by individuals from a neighbouring group, flock or pack. Simulation analysis of similar systems, which account for territorial behaviour but not resource selection (Giuggioli, Potts & Harris 2011; Potts, Harris & Giuggioli 2012, 2013), shows that the resulting territories are not fixed in space. Therefore, including territorial interactions would require that the boundary, $\Omega$, in the ME (eqn 3) were replaced by one that varies in time. Whilst this is not necessary for the caribou population modelled here, as they do not form territories, for many animals, this is an important consideration in linking individual mechanisms to the population patterns. We hope to include this in future studies.

### Acknowledgements

This study was partly funded by NSERC Discovery and Acceleration grants (MAL, JRP). MAL also gratefully acknowledges a Canada Research Chair and a Killam Research Fellowship. This study was partly funded by the Institute for Biodiversity, Ecosystem Science & Sustainability and the Sustainable Development & Strategic Science Division of the Newfoundland & Labrador Department of Environment & Conservation. GBR was supported by a scholarship from the NSERC. Caribou telemetry data were provided by the Newfoundland and Labrador Department of Environment and Conservation. We thank all the government personnel involved in the caribou captures and monitoring over the last four years. We are grateful to Ulrike Schlägel and Jimmy Garnier from the Lewis Lab for helpful discussions regarding modelling and analysis, as well as two reviewers for useful comments that helped improve the manuscript.

## Supporting Information

Additional Supporting Information may be found in the online version of this article.

**Appendix S1.** Implementing eqn (3) from the main text in discrete space.

**Figure S1.** Schematic representation of eqn (1).